\documentclass[conference]{IEEEtran}
\usepackage{geometry}
\usepackage{blindtext}
\usepackage{inputenc}
\usepackage{graphicx}
\usepackage{comment}
\usepackage{float}
\usepackage{graphics} 
\usepackage{mathtools}
\usepackage{wrapfig}
\usepackage{enumitem}
\usepackage[noend]{algpseudocode} 
\usepackage{float}
\usepackage{algorithmicx}
\usepackage{algorithm}
\usepackage{arydshln}
\usepackage{wrapfig,lipsum,booktabs}
\usepackage{mathtools}
\usepackage{amsmath}
\usepackage{amssymb}

\PassOptionsToPackage{bookmarks=false}{hyperref}
\IEEEoverridecommandlockouts
\usepackage{etoolbox}
\usepackage{subcaption}

\begin{document}
	\title{Representable Matrices: Enabling High Accuracy Analog Computation for Inference of DNNs using Memristors}

	\author{\IEEEauthorblockN{Baogang Zhang\IEEEauthorrefmark{1},
			Necati Uysal\IEEEauthorrefmark{1} \thanks{This paper was accepted at Asia and South Pacific Design Automation Conference 2020.}, Deliang Fan\IEEEauthorrefmark{2} and
			Rickard Ewetz\IEEEauthorrefmark{1} \thanks{This research was supported in part by NSF awards CCF-1755825 and CNS-1908471.}}
		\IEEEauthorblockA{\IEEEauthorrefmark{1}University of Central Florida, Orlando, FL, 32816, USA\\
			\IEEEauthorrefmark{2}Arizona State University, Tempe, AZ, 85281, USA\\ 
			baogang.zhang@knights.ucf.edu,
			necati@knights.ucf.edu,
			dfan@asu.edu,
			rickard.ewetz@ucf.edu}}

	\floatname{algorithm}{Algorithm}
	\renewcommand{\algorithmicrequire}{\textbf{Input:}}
	\renewcommand{\algorithmicensure}{\textbf{Output:}}
	\newcommand*\Let[2]{\State #1 $\gets$ #2} 
	\algnewcommand{\LineComment}[1]{\State \(/*\) #1 \(*/\)}
	\algnewcommand{\Cotton}[1]{\hfill\(/*\) #1 \(*/\)}

	\maketitle
	
	\makeatletter
	\begin{abstract}
		Analog computing based on memristor technology is a  promising solution to accelerating the inference phase of deep neural networks (DNNs). A fundamental problem is to map an arbitrary matrix to a memristor crossbar array (MCA) while maximizing the resulting computational accuracy. The state-of-the-art mapping technique is based on a heuristic that only guarantees to produce the correct output for two input vectors. In this paper, a  technique that aims to produce the correct output for every input vector is proposed, which involves specifying the memristor conductance values and a scaling factor realized by the peripheral circuitry. The key insight of the paper is that the conductance matrix realized by  an MCA is only required to be proportional to the target matrix. The selection of the scaling factor between the two regulates the utilization of the programmable memristor conductance range and the representability of the target matrix. Consequently, the scaling factor is set to balance precision  and value range errors. Moreover, a technique of converting conductance values into state variables and vice versa is proposed to handle memristors with non-ideal device characteristics. Compared with the state-of-the-art technique, the proposed mapping results in 4X-9X smaller errors. The improvements translate into that the classification accuracy of a seven-layer convolutional neural network (CNN) on CIFAR-10 is improved from 20.5\% to 71.8\%. 
	\end{abstract}

	\section{Introduction}
	Deep neural networks (DNNs) have in recent  years achieved remarkable results in terms of image, audio, and video recognition~\cite{Deep:2015}. The arising  solution to enable the computationally heavy DNNs to be deployed on edge-devices in the Internet of Things is to leverage memristor-based technology. Memristor crossbar arrays (MCAs) can perform matrix-vector multiplication in the analog domain with orders of magnitude smaller  power and latency than in the digital domain~\cite{Hu:2016,Hu:2018}. Moreover, the use of MCAs allow matrices to be stored in-place, which reduces data fetching and communication costs that fundamentally bounds the performance of any computing system that processes large amounts of data~\cite{ISAAC:2016,Chi:2016,Song:2017}.
	
\begin{figure}
		\centering
		\small
		\includegraphics[width=4cm]{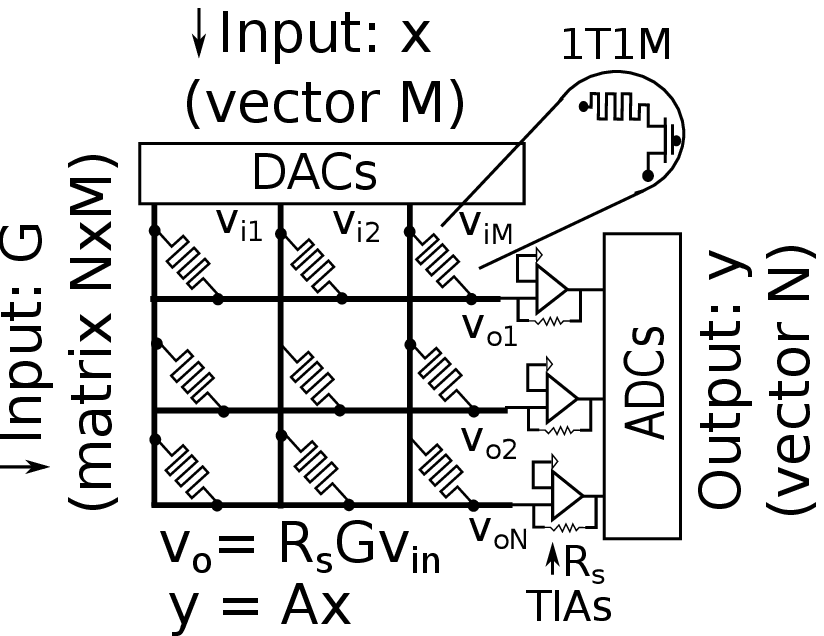} 
		\includegraphics[width=4cm]{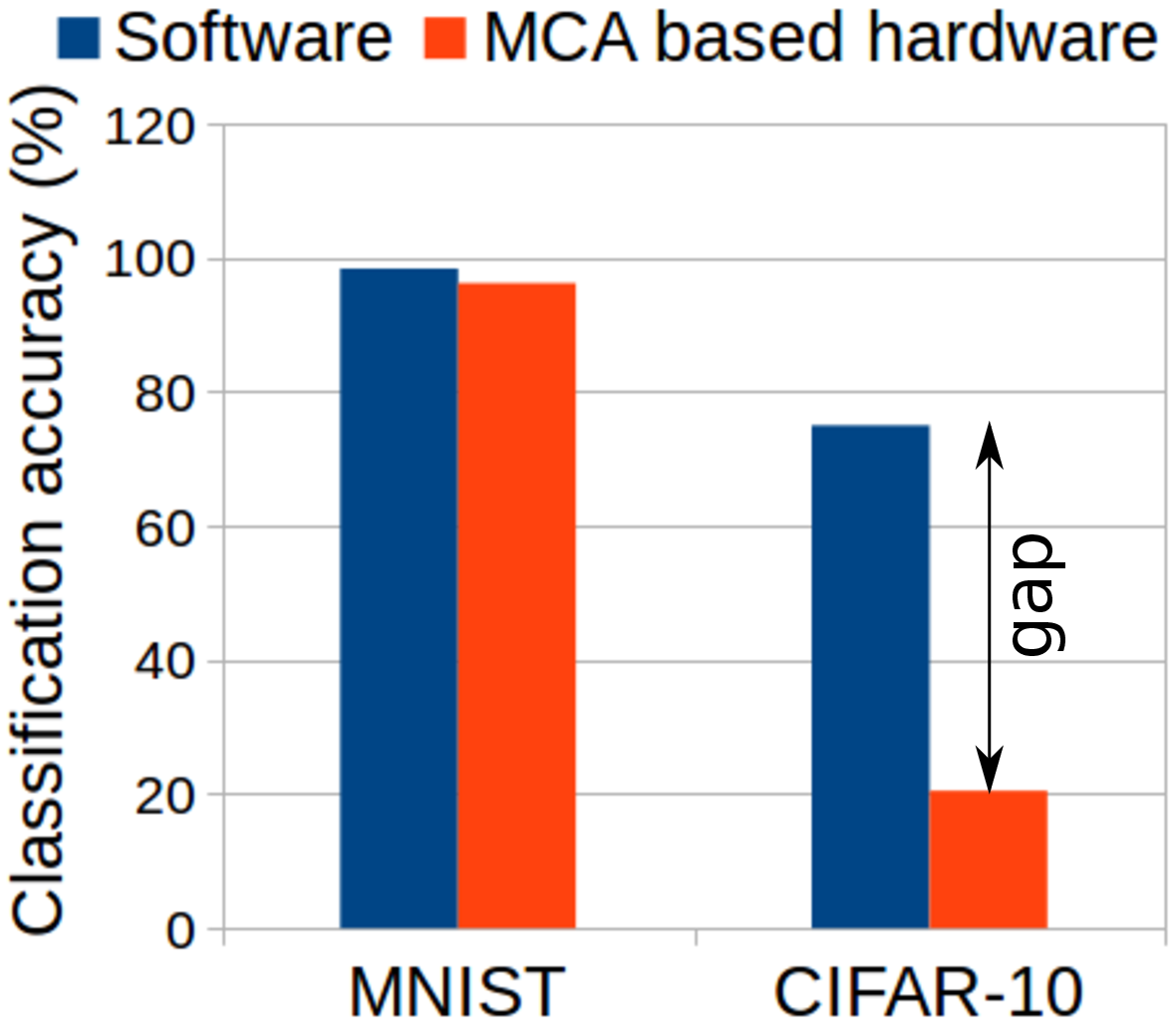}\\ \vspace{-2pt}
		(a) \hspace{4cm} (b)
		\vspace{-3pt}
		\caption{(a) An MCA used for matrix-vector multiplication. (b) Classification accuracy in software and MCA based  hardware on MNIST and CIFAR-10 using the  mapping  in~\cite{Hu:2016}.}
		\label{fig:mca}
		\vspace{-15pt}
	\end{figure}

	Matrix-vector multiplication is performed using an MCA with access transistors by first programming the conductance values of the memristors to realize a conductance matrix $G$, which is illustrated in Figure~\ref{fig:mca}(a)~\cite{Hu:2014}. Next, an input vector of voltages ($v_{in}$) are applied to the vertical columns and a vector of output voltages ($v_{out}$) are measured from the horizontal rows. The inputs are provided to the MCA using digital-to-analog converters (DACs) and the outputs are converted into digital values   using transimpedance amplifiers (TIAs) and analog-to-digital converters (ADCs). The output voltages $v_o$ are equal to $R_sGv_{i}$, where $R_s$ is the feedback resistances of the TIAs~\cite{Hu:2014,Hu:2016}. Next, the output voltages are scaled into digital values. However, if a weight matrix is mapped to an MCA without considering effects as IR-drop, programming errors, and non-ideal device characteristics, the computational accuracy will be degraded into noise~\cite{Hu:2016}. In particular, the accuracy is degraded by the IR-drop across the \mbox{non-zero} input, output and wire resistance in the MCA.
	
	Techniques to map an arbitrary matrix $W$ to an MCA have been studied in~\cite{Hu:2014,Xia:2016,Liu:2014,Hu:2016}. The conductance matrix $G$ realized by a set of memristor conductance values $g$ can be determined analytically using Modified Nodal Analysis (MNA)~\cite{Liu:2014}.
	Next, the effective matrix realized by an MCA ($W^r$) is obtained by scaling $G$ with a factor $(1/\alpha)$, which is realized by the peripheral circuitry. In~\cite{Liu:2014}, the conductance values $g$ were determined by minimizing the square of the Frobenius norm of $(W-W^r)$ using steepest gradient decent. However, the method is unable to consistently converge for arbitrary matrices. In~\cite{Hu:2016}, a technique of tunning conductance values (or state variables  of non-ideal memristor devices) to minimize $||(W - W^r) \cdot v_{cal}||^2$ using Newton's method was proposed, where $v_{cal}$ is a calibration vector. The main limitation of these works is that the scaling factor $\alpha$ was not explicitly optimized.  
	Nevertheless, the technique in~\cite{Hu:2016} enabled a five-layer feed-forward neural network to be mapped to a memristor based platform while achieving software level accuracy on the MNIST dataset. However, when the technique is used to map a seven-layer convolutional neural network (CNN) to a MCA  based platform, the classification accuracy drops from $75.2\%$ to $20.5\%$, which is shown in Figure~\ref{fig:mca}.

	In this paper, a technique is proposed to map an arbitrary matrix $W$ into a scaling  factor $\alpha$ and memristor state variables $s$.
	The main innovations of the paper are summarized, as follows:
	\vspace{-4pt}
	\begin{itemize}
		\item The problem of specifying ($s$, $\alpha$) is converted into a problem of specifying ($g$, $\alpha$). The technique is based on replacing each series connected memristor and access transistor  with an ideal conductor. After the conductance values $g$ have been determined, Newton's method is used to obtain the equivalent state variables $s$. 
		
		\item The memristor conductance values $g$ are only required to be specified to realize a conductance matrix $G$ that is proportional to the matrix $W$. Next, the conductance matrix is effectively scaled with $1/\alpha$ such that $W$ is effectively realized.
		Nevertheless, the utilization of the memristor conductance range is also regulated by the scaling factor $\alpha$. If $\alpha$ is set too small, the errors will be dominated by the limited \emph{precision} of the memristors. If $\alpha$ is set too large, the errors will be dominated by \emph{value errors} introduced by IR-drop. In particular, it is impossible to represent small and large  values in the far-end of an MCA because of IR-drop. We refer to the insights of the matrices that can be realized at different locations in an MCA as defining the space of representable matrices. Consequently, $\alpha$ is specified while balancing precision and value range errors.

		\item Given $\alpha$, the conductance values $g$ are specified minimizing $||(W - W^r) ||^2_F$ using steepest gradient decent, where $||(W - W^r) ||^2$ is called the total errors. After it is impossible to further reduce the total errors, the matrix $W$ is updated into a new target matrix $W^t$ in order to ensure that $||(W - W^r) \cdot v_{cal}||^2=0$, where $v_{cal}$ is a calibration vector  selected from the input vector space. .

		\item The experimental results  show that the proposed mapping technique results matrix-vector multiplication with $4X$-$9X$ higher computational accuracy  compared with in~\cite{Hu:2016}. The improvements translate into  that the classification accuracy of a seven-layer CNN is improved from $20.5\%$ to $71.8\%$ on CIFAR-10, which is close to the software accuracy of $75.2\%$. 
	\end{itemize}
	
	\section{Preliminaries}
	\label{sec:pre}
	\vspace{-3pt}
	\subsection{Circuit model of MCAs~\cite{Liu:2014,Hu:2016}}
\begin{figure}[h]
		\centering
		\small
		\includegraphics[width=8cm]{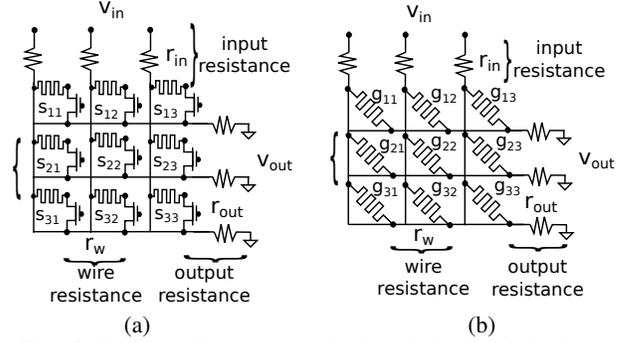}\\
		\vspace{1pt}
		(a) \hspace{4.0cm} (b)
		\vspace{-5pt}
		\caption{MCA with (a) non-ideal  and (b) ideal devices.}
		\label{fig:mvm}
		\vspace{-12pt}
	\end{figure}

	Figure~\ref{fig:mvm} shows two circuit models for the MCAs in Figure~\ref{fig:mca}. In Figure~\ref{fig:mvm}(a), the memristors and the access transistors are modeled using non-linear equations. The current $i_m(s,v_m)$ through a memristor is a \mbox{non-linear} function of the state variable $s$ and the voltage $v_m$ across the device. The current through each access transistor $i_t(v_s,v_d,v_g)$ is a \mbox{non-linear} function of the source, drain, and gate voltage. A simplified circuit model of the MCA when both the memristors and the access transistors are treated as ideal devices is shown in Figure~\ref{fig:mvm}(b), which allows them to be replaced with a single ideal memristor $g$ (or an conductor with lower and upper bounds).
	
	\subsection{Matrix realized by an MCA}
	The matrix $W^r$ realized by an MCA is  a function of $g$ and $\alpha$, i.e., $W^r = f(g,\alpha)$. 
	For the  circuit model in Figure~\ref{fig:mvm}(b), the realized matrix $W^r$ can be obtained  analytically, as follows:
	\begin{align}
		W^r = G/\alpha \label{eq:ana}
	\end{align}
	where $G$ is the conductance matrix realized by the resistive network and  and $\alpha$ is an arbitrary scaling factor realized by the peripheral circuitry. The conductance matrix $G$ is a non-linear function of the memristor conductance values $g$. $G$ is obtained by formulating a system on linear equations that capture the resistive network using MNA, as follows:
	\begin{align}
		Y(g) \begin{bmatrix} v \\ v_{dac} \end{bmatrix}  = \begin{bmatrix} 0 \\ v_{in} \end{bmatrix}, \label{eq:mna}
	\end{align}
	where $Y(g)$ is a matrix with dimensions $(2NM +M)$x$(2NM +M)$ that is a function of $g$. $M$ and $N$ are the number of inputs and outputs, respectively. 
	$v$ and $v_{dac}$ are respectively the node voltages of the MCA and DACs; $v_{in}$ is the input voltages. Next, $G$ is obtained analytically, as follows:
	\begin{align}
		G = SY^{-1}(g)B, \label{eq:ana2} 
	\end{align}
	where $B = [0, 0, I]^T$ and $I$ is an $M$x$M$ identity matrix. $S$ is an $N$x$(2 \cdot N \cdot M + M)$ matrix that selects the output voltages from $Y^{-1}B$.

	\begin{figure*}[h] 
		\begin{centering}
			\includegraphics[width=16cm]{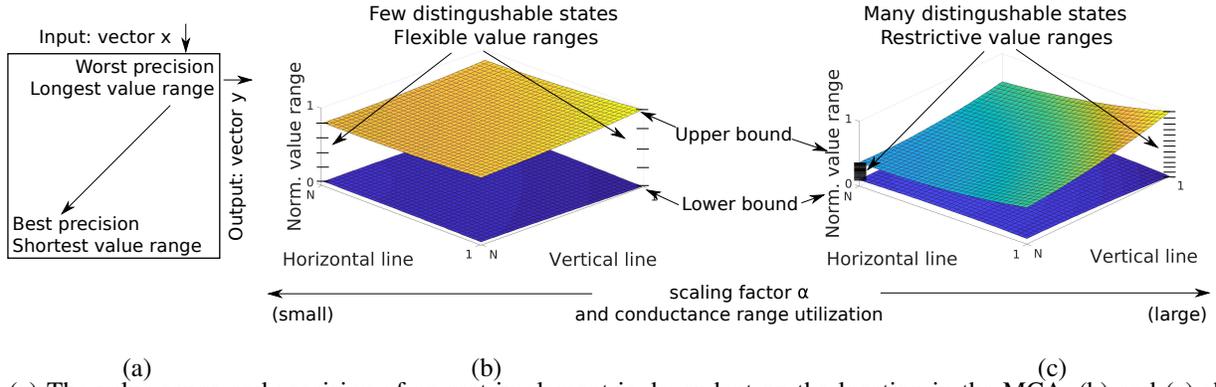}\\ 
			\hspace{-0.5cm} (a) \hspace{4cm} (b)\hspace{7cm} (c) \hspace{4cm}\\ \vspace{-7pt}
			\caption{(a) The value range and precision of an matrix element is  dependent on the location in the MCA. (b) and (c) shows a trade-off between the value ranges and the precision based on the scaling factor $\alpha$.}
			\label{fig:location}
		\end{centering}
	\vspace{-10pt}
	\end{figure*}
	
	\subsection{Problem formulation}
	\label{sec:problem}
	This paper considers the problem of specifying  the memristor state variables $s$ and the scaling factor $\alpha$ that maximizes the computational accuracy when performing matrix-vector multiplication using an MCA. If the memristor devices are ideal, the problem consists of specifying the conductance values $g$ and the scaling factor $\alpha$. The computational accuracy is evaluated while accounting for that memristors only can be programmed to a limited number of  distinguishable states~\cite{Alibart:2012,Hu:2018}. In the experimental results, the  impact of the proposed mapping technique is also evaluated in terms of classification accuracy when DNNs trained in software are mapped to an MCA based platform for inference. 
	
	The problem is approached by first converting problem of specifying state variables and a scaling factor ($s$, $\alpha$) into a problem of specifying memristor conductance values and a  scaling  factor ($g$, $\alpha$). This enables $g$ to be specified while minimizing $||(W-W^r)||^2$ or $||(W-W^r) \cdot v_{cal} ||$. Lastly, the conductance values $g$ are converted back into state variables $s$.

	\section{Previous work}
	\label{sec:previous}
	\subsection{Specification of conductance values $g$ in~\cite{Liu:2014}}
	\label{sec:review1}
	In~\cite{Liu:2014}, $\alpha$ was fixed and the conductance values $g$ were determined by formulating an optimization problem, where the square of the Frobenius norm of $(W-W^r)$ was minimized, as follows:
	\begin{align}
		\min F(g,\alpha) = ||W - W^r ||^2  =  \sum_{i=1}^N\sum_{j=1}^M (w_{ij} - w_{ij}^r)^2, \label{eq:error}
	\end{align}
	where $||.||^2$ is the square of the Frobenius norm. $w_{ij}$ and $w^r_{ij}$ are the elements in row $i$ and column $j$ in the weight matrix $W$ and the realized matrix $W^r$, respectively. The function $F$ is minimized using steepest gradient decent, as follows: 
	\begin{align}
		g_{k+1} &= g_k + t  \triangledown F(g) \nonumber \\
		&= g_k + t \cdot   \sum_{i=1}^N \sum_{j=1}^M 2 \cdot (w_{ij} - w_{ij}^r) \cdot \frac{\partial w^r_{ij}}{\partial g}, \label{eq:step}
	\end{align}
	where $\frac{\partial w^r_{ij}}{\partial g}$ is the derivative of $w_{ij}$ with respect to $g$. $t$ is the step size, which is determined using a linear search. $g_0$ is equal to $W$ linearly mapped into the memristor conductance range. Iterative tuning is performed to compensate for IR drop in the MCA. 
	The main limitations is that the method is unable to consistently converge to solutions with high accuracy because $\alpha$ was fixed.

	\vspace{-4pt}
	\subsection{Specification of state variables $s$ in \cite{Hu:2016}}
	\label{sec:review2}
	In~\cite{Hu:2016}, a technique of specifying the state variables of the memristors $s$ was  proposed. The matrix $W$ is first linearly mapped into the programmable memristor conductance range to obtain an ideal conductance matrix $G_{ideal}$. The ideal current ($i_{ideal}$) through each memristor device is obtained using $G_{ideal}$ and an input calibration vector $v_{cal}$, which also implicitly defines the scaling factor $\alpha$. Next, MNA is used to formulate a system of $(4NM + 2N + 2M)$x$(4N M + 2N + 2M)$ equations to capture the circuit model in Figure~\ref{fig:mvm}(a). $NM$ of the equations are used to force the currents through each memristor to be equal to $i_{ideal}$ and the remaining equations are used to capture the behavior of the circuit. The state variables $s$ are determined by solving the system of equations using Newton's method. If Newton's algorithm does not converge, $i_{ideal}$ is updated to ensure that the full programmable conductance range was utilized. The limitation is that only the zero input vector ($\bar{0}$) and the calibration vector ($v_{cal}$) are guaranteed to produce the correct output, i.e., $||(W -W^r) \cdot \bar{0}||^2 = 0$ and  $||(W -W^r) \cdot v_{cal}||^2 = 0$.

	\section{Space of Representable Matrices}
	\label{sec:rep}
	In this section, we define the space of matrices that are representable using an MCA and analyse the impact of the scaling factor $\alpha$. The observations motivates our proposed mapping technique. 
	
	The \emph{value range} and \emph{precision} for each matrix element is dependent on the \emph{location} in the MCA and the scaling factor $\alpha$, which is illustrated in Figure~\ref{fig:location}. Based on Eq~(\ref{eq:ana}) and Eq~(\ref{eq:ana2}), the scaling factor $\alpha$ directly regulates the utilization of the programmable conductance range, i.e., a larger $\alpha$ implies a utilization of larger conductance values. The value range for a matrix element consists of a lower and upper bound on the value that can be realized. The upper bound mainly stems from IR-drop. The lower bound stems from that currents may flow from an vertical line $i$ to an horizontal line $j$ even if the memristor device connecting vertical line $i$ to horizontal line $j$ is set to be non-conductive (maximum resistance), i.e., the current would flow on paths in the MCA containing more than one memristor. The length of the value range is the longest in the top-right corner and the shortest in the bottom-left corner of an MCA. Moreover, there is an equal number of distinguishable states between every lower and upper bound. The number of states is dependent on the accuracy of the closed loop programming and the selected utilization of the programmable conductance range. Consequently, the worst (best) precision is obtained for value range's with the longest (shortest) length, which is illustrated in Figure~\ref{fig:location}(a). Moreover, by reducing the utilization of the conductance range, every value range becomes more flexible at the expense of worse precision because the number of distinguishable states within each value range is reduced, which is illustrated in (b-c) of Figure~\ref{fig:location}. The explanation is that utilization of  high conductive increases the currents on the paths with multiple memristors and IR-drop. 
	
\begin{figure}[h]
		\centering
		\small
		\includegraphics[width=4cm]{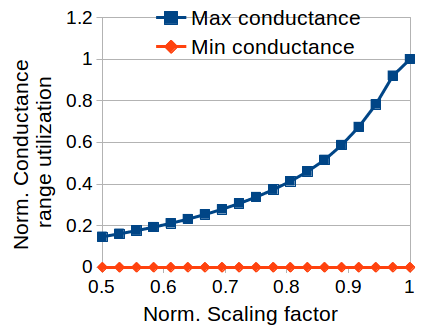}
		\includegraphics[width=4cm]{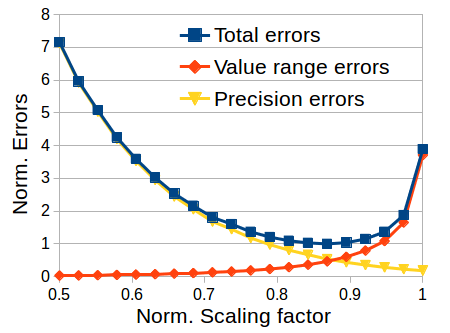}\\
		(a) \hspace{3.5cm} (b)\\
		\caption{(a) Memristor conductance range utilization vs. scaling factor $\alpha$. (b) Total errors, value range errors, precision errors vs. scaling factor $\alpha$. The results are obtained for an MCA with dimensions $64$x$64$.}
		\label{fig:trade}
	\end{figure}

	The described observations directly explain the space of matrices that are representable using an MCA, i.e., every matrix can be realized using an MCA but the computational accuracy depends on the sum of the value range errors and the precision errors (called total errors). Value range errors occur  when matrix elements are attempted to be realized outside their respective value ranges. The precision errors depend profoundly on how large portion of the programmable memristor conductance range is utilized. Consequently, a critical problem is to specify the scaling factor $\alpha$ to balance the value range and the precision errors, which is shown in Figure~\ref{fig:trade}.

	In Figure~\ref{fig:trade}(a), it is shown that the utilization of the programmable memristor conductance range is dependent on $\alpha$. In Figure~\ref{fig:trade}(b), the trade-off between value range errors and precision errors is shown based on $\alpha$. If $\alpha$ is selected too small, the value range errors will neglectable but large precision errors will be introduced. If $\alpha$ is selected to large, the precision errors will be neglectable but large value range errors will be introduced. Hence, $\alpha$ should be selected so only a few values are slightly outside the value ranges such that the algorithm in~\cite{Liu:2014} can be used to specify the conductance values by minimizing $||W-W^r||$.

	\section{Proposed methodology}
\begin{figure}[h]
		\centering
		\small
		\includegraphics[width=7cm]{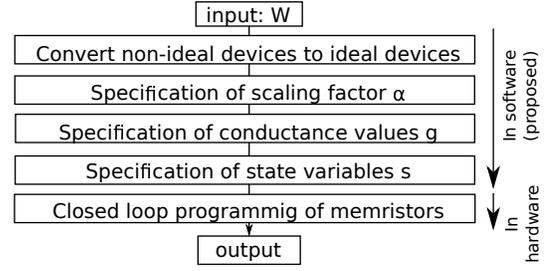}\\

		\caption{Proposed flow for mapping $W$ to an MCA.}
		\label{fig:flow}
		\vspace{-12pt}
	\end{figure}

	We propose a five step flow to map an arbitrary target matrix $W$ to an MCA which is shown in Figure~\ref{fig:flow}. The first step consists of replacing each non-ideal memristor and access transistor with an equivalent ideal memristor. As mentioned earlier,  the conversion is performed to allow $W^r$ to be computed using Eq~(\ref{eq:ana}). The details are provided in Section~\ref{sec:step1}. The second step is to determine the scaling factor $\alpha$ that minimizes the total errors $||W-W^r||^2$, which is  explained in Section~\ref{sec:step2}. The third step is to specify the conductance values $g$ that minimize the total errors while guaranteeing that, $||(W-W^r) \cdot v_{cal}||$, is close to zero, which is outlined in Section~\ref{sec:step3}. The motivation is to leverage the known properties of the input vector space. Fourth, the state variables $s$ of the non-ideal memristors are determined by solving a system of non-linear equations using Newton's method, as explained in Section~\ref{sec:step4}. Lastly, closed loop programming is applied to program the memristors on-chip to the desired states $s$ using the techniques in~\cite{Alibart:2012,Hu:2018}. 
	
	\label{sec:method}
	
	\subsection{Convert non-ideal devices to ideal devices}
	\label{sec:step1}
	The first step is to convert the non-ideal memristors and access transistor to  ideal memristor with a conductance of $g$, i.e., converting the circuit in  Figure~\ref{fig:mvm}(a) to the circuit in Figure~\ref{fig:mvm}(b). $g$ is bounded within $[g_{min}, g_{max}]$, where $g_{min}$ and $g_{max}$ are respectively the conservatively estimated minimum and maximum conductance of the series connection of each memristor and access transistor. These conductance values are  different from the programmable memristor conductance range because the estimated conductance of the access transistor is included. %

	\subsection{Specification of scaling factor $\alpha$}
	\label{sec:step2}
	In this section, a technique of specifying the  scaling factor $\alpha_{opt}$ that minimizes the 
	total errors $||(W-W^r)||^2$ is proposed. 
	
	The method used in this paper is based on first guessing an scaling factor $\alpha_0$. Given $\alpha_0$, the technique in~\cite{Liu:2014} is utilized to specify the memristor conductance values $g$ by minimizing $||W-W^r||$. Next, $\alpha_k$ is updated to $\alpha_{k+1}$ based on the relation between the value range errors and the conversion errors to minimize the total errors. 
	If the value range errors are larger than the precision errors, $\alpha_k$ is updated to $\alpha_{k+1} = \alpha_k \cdot (1-\beta)$. If the precision errors are larger than the value range errors, $\alpha_{k+1} = \alpha_{k} \cdot (1 + \beta)$. (Experimentally we have observed that the total errors are close to the minimum when the value range errors are equal to the precision errors.) The process is repeated until no further improvements in terms of total errors are achieved over $t$ iterations. The parameters $\beta$ and $t$ are set to  balance a trade-off between errors and run-time.  
	
	Specifically, the total errors are obtained by first quantizing $g$ (based on the bit-accuracy of the closed loop programming). Next, $W^r$ is obtained using Eq~(\ref{eq:ana}) and the total errors are computed as $||W -W^r||^2$. The value range errors are obtained as $||W-W^r||^2$ without first quantizing the conductance values $g$. The precision errors are set to the difference between the total errors and the value range errors.

	\subsection{Specification of conductance values $g$}
	\label{sec:step3}
	In this section, the conductance values $g$ are specified given $W$, $\alpha_{opt}$, and a calibration vector $v_{cal}$ from the input vector space. The objective is to minimize $||W-W^r||^2$ while guaranteeing that $||(W-W^r) \cdot v_{cal}|| =0 $. 
	The motivation for minimizing $||(W-W^r) \cdot v_{cal}||$ is to exploit that the outputs from neurons in DNNs are non-negative due to the activation functions.

	This step is performed by updating $W$ to a new target matrix $W^t$. Next, given $W^t$ and $\alpha_{opt}$ the conductance values $g$ are specified using the approach based on minimizing $||W^t-W^r||^2$ using the method in~\cite{Liu:2014}. Unfortunately, it is impossible eliminate the errors by updating elements in $W$ that are realized too small (or too large) because the corresponding memristors are already tuned to the lower (or upper) bound of the programmable memristor conductance range. Hence, the errors are distributed to the matrix elements in the same row  to ensure that $||(W-W^r)\cdot v_{cal}||=0$. 
	
	Let $R = (W^r-W)$ be the difference between the realized matrix $W^r$ and the matrix $W$. Next, let $r$ be a vector containing the sum of the elements in each row of $R$ and let $c$ be a vector containing the number of memristors with a conductance not equal to $g_{min}$ in each row of the MCA. Next, let $u$ be equal to $r$ element-wise divided by $c$. Subsequently, $W$ is updated to $W^t$ by adding $u(i)$  to each element in row $i$ where the conductance of the corresponding memristor is not equal to $g_{min}$. Next, $W^t$ is mapped into conductance values using the technique in~\cite{Liu:2014}.

	\subsection{Specification of state variables $s$}
	\label{sec:step4}
\begin{figure}[h]
		\centering
		\small
		\vspace{-5pt}
		\includegraphics[width=5.5cm]{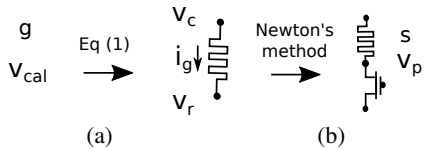}\\
		(a) \hspace{2.5 cm} (b) 
		\caption{Specification of $s$ from $g$ and $v_{cal}$.}
		\label{fig:nonideal}
	\end{figure}

	In this section, the state variables of the memristors are determined from the ideal conductance values $g$ and a calibration signal $v_{cal}$, which is illustrated in Figure~\ref{fig:nonideal}. First, the node voltages $v_c$ and $v_r$ and the currents through the conductors $i_g$ are computed using $g$ and $v_{cal}$, which is shown in Figure~\ref{fig:nonideal}(a). Next, the state variables $s$ are computed using $v_c$, $v_r$, and $i_g$, which is illustrated in Figure~\ref{fig:nonideal}(b).
	
	\textbf{Computation of $v_c$, $v_r$, $i_g$:} First, the node voltages $v_c$ and $v_r$ in the MCA are computed with respect to a calibration signal $v_{cal}$ by solving Eq~(\ref{eq:ana}) with $v_{in}$=$v_{cal}$. We use $v_{cal}$=$v_{max}/2$ in our implementation ($v_{cal}$ can also be set based on prior knowledge of the input vectors of a specific application). Next, the currents though each conductor $g$ is obtained using $i_g = g\cdot (v_c - v_r)$.
	
	\textbf{Computation of state variables $s$:} The state variables $s$ are found by solving a non-linear system of two equations, as follows: 
	\begin{align}
		X = \begin{bmatrix} s \\  v_p \end{bmatrix}, \quad F(X) = i_g - \begin{bmatrix} i_m(s,v_c\text{-}v_p) \\ i_t(v_p,v_r,v_g) \end{bmatrix},  
	\end{align}
	where $v_p$ is the node voltages between the memristors and the access transistors. Next, Newton's method is used to solve for $F(X) = 0$, as follows:
	\begin{align}
		X_{k+1} = X_k - J^{-1}F(X_k),  \label{eq:new}
	\end{align}
	where $J^{-1}$ is the inverse of the Jacobian of $F$. Note that Newtons method can be applied independently for each memristor and access transistor pair.

	\section{Experimental evaluation}
	\label{sec:results}
	The experimental results are obtained using a quad core 3.4 GHz Linux machine with 32GB of memory. The proposed techniques are implemented in MATLAB. The default MCA in the evaluation has dimensions $128$x$128$, a wire resistance $r_w$=$1\Omega$, and both the input and output resistance are $100 \Omega$. The programmable memristor conductance range is $2k \Omega$ to $3M \Omega$~\cite{Hu:2016}. We use the same non-ideal device models for the memristors and access transistors as in~\cite{Hu:2016}, which is  available in~\cite{Strachan:2013}. The bit-accuracy for the memristors is set to $8$ bits. 
	The programming errors are modeled using quantization, where it is assumed that the distinguishable states in memristor conductance range (or state space) are equidistant. The maximum input voltage is set to $0.2V$. The determined state variables $s$ and scaling factors $\alpha$ are evaluated using circuit simulation with SPICE accuracy using the circuit model in Figure~\ref{fig:mvm}(a). In Section~\ref{sec:mvm}, the proposed mapping is evaluated in terms of matrix-vector multiplication. In Section~\ref{sec:results2}, the proposed mapping technique is evaluated in an DNN application. We compare our results with the technique proposed in~\cite{Hu:2016}. No direct comparison is provided with in~\cite{Liu:2014}, since that work considered a subproblem of our problem formulations.

	\subsection{Evaluation of matrix-vector multiplication}
	\label{sec:mvm}
\begin{figure}[h]
		\centering
		\includegraphics[width=4.2cm]{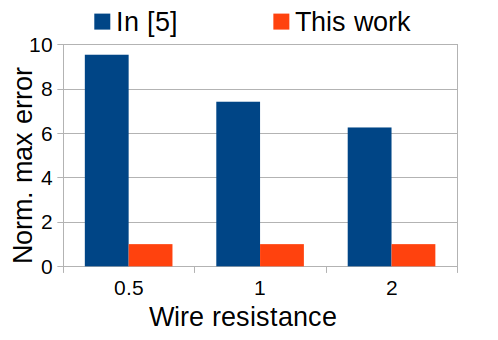}
		\includegraphics[width=4.2cm]{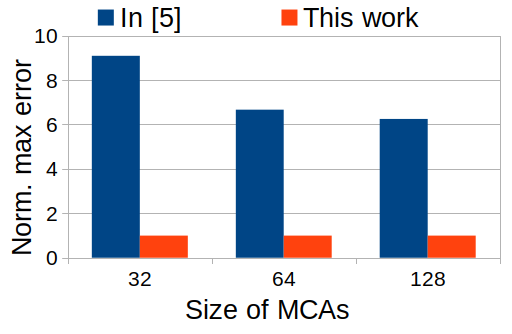}\\
		(a) \hspace{3.5cm} (b) \\
		\includegraphics[width=4.2cm]{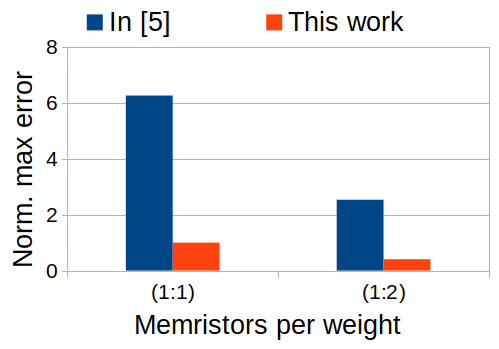}
		\includegraphics[width=4.2cm]{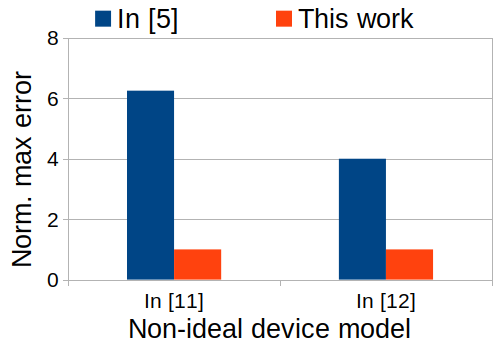}\\
		(c) \hspace{3.5cm} (d)\\
		\vspace{-5pt}
		\caption{Comparison with  \cite{Hu:2016} using different (a) wire resistance, (b) crossbar sizes, (c) number of memristors per weight, and (d) non-ideal device models.}
		\vspace{-6pt}
		\label{fig:comp}
	\end{figure}

	In this section, we compare the proposed mapping technique with the state-of-the-art mapping technique in~\cite{Hu:2016} using full analog simulation using state variables $s$. The evaluation is performed with respect to the maximum output error for various input vectors and weight matrices. In Figure~\ref{fig:comp}, it is demonstrated that the proposed mapping results in $4X$ to $9X$ smaller errors based on the wire resistance, crossbar size, number of memristors used per matrix element/weight, and memristor device model. It is not surprising that significant smaller maximum output errors are obtained because \cite{Hu:2016} is a heuristic and the proposed mapping technique specifies both $g$ and $\alpha$ by leveraging the insights provided by the space of representable matrices. Since the benefits are obtained using only parameter optimization, the power and area is expected to be extremely similar to in~\cite{Hu:2016,Hu:2018}, i.e., the benefits are obtained with no overhead. In general, we find that when square MCAs of size $32$/$64$/$128$ are used, $64/60/48\%$ of the programmable memristor conductance range is utilized. The average run-time is $0.5$/$1.5$/$5$ min per matrix with a dimension of $32$/$64$/$128$, respectively. Note that for MCAs of dimension 128x128, only a few $\alpha$ values were evaluated in order to control the run-time.

	\subsection{Evaluation of DNN applications}
	\label{sec:results2}
\begin{figure}[h]
		\centering
		\vspace{-5pt}
		\begin{minipage}[t]{.4\columnwidth}
			\vspace{0pt}
			\includegraphics[height=2.5cm]{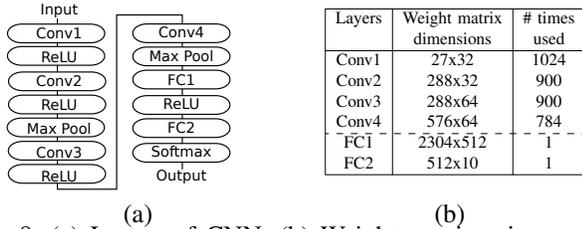} 
		\end{minipage}\qquad
		\begin{minipage}[t]{.4\columnwidth}
			\vspace{2pt}
			\resizebox{\columnwidth}{!}{%
\begin{tabular}{|c|c|c|}
\hline
Layers  & Weight matrix & \# times  \\
 & dimensions &  used \\\hline
Conv1 & 27x32 & 1024 \\
Conv2 & 288x32  & 900 \\
Conv3 & 288x64  & 900 \\
Conv4 & 576x64  & 784 \\ \cdashline{1-3}
FC1 & 2304x512 & 1\\ 
FC2 & 512x10 & 1\\ \hline
\end{tabular}
}
		\end{minipage} \\
		\vspace{5pt}
		(a) \hspace{3.5cm} (b) \vspace{-8pt}
		\caption{(a) Layers of CNN. (b) Weight matrices in convolutional (Conv) and fully-connected layers (FC). }
		\vspace{-5pt}
		\label{fig:prop}
	\end{figure}

	In this section, the proposed mapping and the method in~\cite{Hu:2016} are used to map DNNs trained in software using GPUs to MCA based platforms for inference. The networks are trained using Keras combined with TensorFlow. In particular, we evaluate a four-layer feed-forward network trained on the MNIST dataset and a seven-layer CNN trained on the CIFAR-10 dataset. The feed-forward network has dimensions $784$x$500$x$300$x$10$ and the properties of the CNN is shown in Figure~\ref{fig:prop}.  
	
	The feed-forward network is mapped to an MCA based platform by partitioning each weight matrix onto a grid of $128$x$128$ MCAs. The CNN is mapped to an MCA based platform using the kernel mapping in~\cite{Song:2017}, where each convolutional layer and fully-connected layer is partitioned onto a grid of $128$x$128$ MCAs. The default settings for the MCA are used with one memristor per weight.  The classification accuracy is computed using one thousand randomly selected input images and SPICE level circuit simulation.

	\begin{figure}[h]
		\centering
		\vspace{-2pt}
		\includegraphics[width=4.2cm]{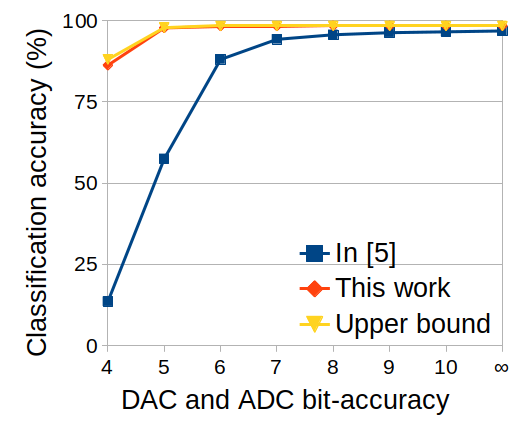} 
		\includegraphics[width=4.2cm]{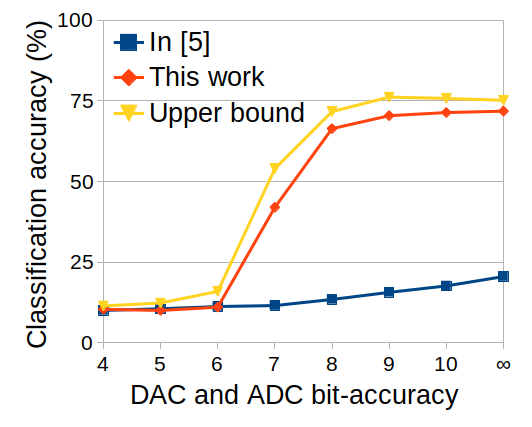}\\
		(a) \hspace{3.5cm} (b)\\ \vspace{-4pt}
		\caption{Classification accuracy for different DAC/ADC bit-accuracies   (a) MNIST (b) CIFAR-10.}
		\label{fig:results}
		\vspace{-5pt}
	\end{figure}  
	
	In Figure~\ref{fig:results}, the  classification accuracy achieved in  MCA based hardware is shown for DACs and ADCs with various bit accuracies on the MNIST and CIFAR-10 datasets. We also plot the upper bound on the classification accuracy that can be achieved using DACs and ADCs, i.e., errors are only introduced by the DACs and ADCs. The DACs and ADCs use a fixed and dynamic reference voltage, respectively. In Figure~\ref{fig:results}(a), it can be observed that when no DAC/ADC quantization is performed (indicated with $\infty$) the classification accuracy on the MNIST dataset using the mapping this work and in~\cite{Hu:2016} is $98.3\%$ and $96.3\%$, respectively. The upper bound or software accuracy is $98.4\%$. Moreover, the proposed mapping technique follows the upper bound closely and outperforms the mapping in~\cite{Hu:2016} when DACs and ADCs with smaller bit-accuracies are used. In  Figure~\ref{fig:results}(b), it can be observed that the proposed mapping achieves a classification accuracy of $71.8\%$, whereas the mapping in~\cite{Hu:2016} results in an accuracy of $20.5\%$ on CIFAR-10, which is close to the  software classification accuracy of $75.2\%$.
	Moreover, the proposed technique follows the upper bound on CIFAR-10 closely for DACs and ADCs with different bit-accuracies. 
	It is easy to understand that the proposed mapping technique outperforms the method in~\cite{Hu:2018} because each matrix-vector multiplication is performed with $6X$ smaller errors.

	\begin{figure}[h]
		\vspace{-5pt}
		\centering
		\begin{minipage}[t]{.4\columnwidth}
			\includegraphics[width=4.2cm]{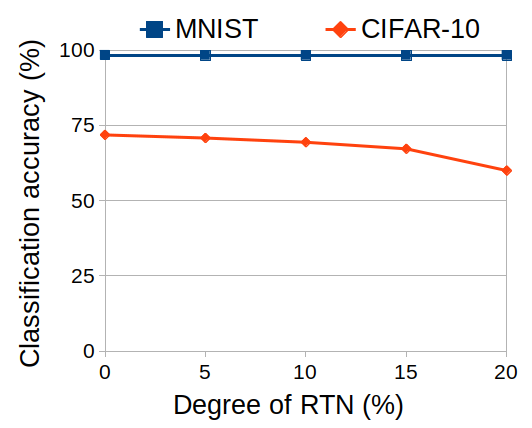} 
		\end{minipage}\qquad
		\begin{minipage}[t]{.4\columnwidth}
			\vspace{-70pt}
			\small
			\resizebox{1.0\columnwidth}{!}{%
				\begin{tabular}{|c|r|r|}
					\hline
					Work & \multicolumn{2}{c|}{Mapping time} \\ 
					& \multicolumn{2}{c|}{(h)} \\
					\cline{2-3}
					& \multicolumn{1}{c|}{MNIST} & \multicolumn{1}{c|}{CIFAR-10} \\ \hline
					In~\cite{Hu:2016} & 0.28 & 0.61 \\
					This work  & 6.17 & 0.93 \\ \hline
				\end{tabular}
			}
		\end{minipage}
		\vspace{3pt}
		\hspace{.65cm}(a) \hspace{3.2cm} (b)\\ \vspace{-5pt}
		\caption{(a) Classification accuracy with RTN. (b) Run-time of mapping DNNs to MCAs.}
		 \vspace{-10pt}
		\label{fig:rr}
	\end{figure}
	
	When  moderate to severe (up to $20\%$) random telegraph noise is included~\cite{Strachan:2015}, the classification accuracy is gracefully reduced (not impacted) on CIFAR-10 (MNIST), which is shown in Figure~\ref{fig:rr}(a).   In Figure~\ref{fig:rr}(b), we show the run-time of our framework and the  techniques in~\cite{Hu:2016}. To limit the run-time of mapping the weight matrices in the CNN to state variables, we used the technique in~\cite{Hu:2016} to map FC1, which is not sensitive to errors. The run-time reported in the table is based on our implementation of~\cite{Hu:2016}, where   performance was prioritized over run-time in the implementation.
	
	\section{Summary and future work}
	\label{sec:summary}
	In this paper, a technique for mapping arbitrary weight matrices to MCAs is proposed. The technique improves the computational accuracy of the state-of-the-art with $4X$ to $9X$ and achieves close to software level accuracy on CIFAR-10 dataset when a CNN trained in software is mapped to an MCA based platform for inference. We plan to reduce the run-time of the algorithm in the future. 
	
	\begin{scriptsize}

	\end{scriptsize}

\begin{thebibliography}{10}

\bibitem{Alibart:2012}
F.~Alibart, L.~Gao, B.~D. Hoskins, and D.~B. Strukov.
\newblock High precision tuning of state for memristive devices by adaptable
  variation-tolerant algorithm.
\newblock {\em Nanotechnology}, 23(7):075201, 2012.

\bibitem{Chi:2016}
P.~Chi et~al.
\newblock {PRIME:} a novel processing-in-memory architecture for neural network
  computation in reram-based main memory.
\newblock ISCA'16, pages 27--39, 2016.

\bibitem{Hu:2014}
M.~Hu et~al.
\newblock Memristor crossbar-based neuromorphic computing system: A case study.
\newblock {\em IEEE Transactions on Neural Networks and Learning Systems},
  25:1864--1878, 2014.

\bibitem{Hu:2018}
M.~Hu et~al.
\newblock Memristor-based analog computation and neural network classification
  with a {DPE}.
\newblock {\em Adv. Materials}, 30, 2018.

\bibitem{Hu:2016}
M.~Hu, J.~P. Strachan, Z.~Li, E.~M. Grafals, N.~Davila, C.~Graves, S.~Lam,
  N.~Ge, J.~J. Yang, and R.~S. Williams.
\newblock Dot-product engine for neuromorphic computing: Programming 1{T}1{M}
  crossbar to accelerate matrix-vector multiplication.
\newblock DAC'16, pages 1--6, 2016.

\bibitem{Deep:2015}
Y.~LeCun, Y.~Bengio, and G.~Hinton.
\newblock Deep learning.
\newblock In {\em Nature}, pages 436--444, 2015.

\bibitem{Liu:2014}
B.~Liu et~al.
\newblock Reduction and {IR}-drop compensations techniques for reliable
  neuromorphic computing systems.
\newblock ICCAD'2014, pages 63--70, 2014.

\bibitem{ISAAC:2016}
A.~Shafiee, A.~Nag, N.~Muralimanohar, R.~Balasubramonian, J.~P. Strachan,
  M.~Hu, R.~S. Williams, and V.~Srikumar.
\newblock {ISAAC:} a convolutional neural network accelerator with in-situ
  analog arithmetic in crossbars.
\newblock ISCA'16, pages 14--26, 2016.

\bibitem{Song:2017}
L.~Song, X.~Qian, H.~Li, and Y.~Chen.
\newblock Pipelayer: A pipelined reram-based accelerator for deep learning.
\newblock HPCA'17, pages 541--552, 2017.

\bibitem{Strachan:2015}
J.~P. Strachan.
\newblock {DPE}: Exploring high efficiency analog multiplication with memristor
  arrays.
\newblock In {\em Int. Conf.on Rebooting Computing}, 2015.

\bibitem{Strachan:2013}
J.~P. Strachan, A.~C. Torrezan, F.~Miao, M.~D. Pickett, J.~J. Yang, W.~Yi,
  G.~Medeiros-Ribeiro, and R.~S. Williams.
\newblock State dynamics and modeling of tantalum oxide memristors.
\newblock {\em IEEE Transactions on Electron Devices}, 60(7):2194--2202, 2013.

\bibitem{Xia:2016}
L.~Xia et~al.
\newblock Technological exploration of {RRAM} crossbar array for matrix-vector
  multiplication.
\newblock {\em Journal of Computer Science and Technology}, 31(1):3--19, 2016.

\end{thebibliography}
\end{document}